# Secure Satellite Communication Systems Design with Individual Secrecy Rate Constraints

Jiang Lei[†], Zhu Han[‡], M. A. Vázquez-Castro[†], and Are Hjørungnes[*]

[†]Dept. Telecom. and Systems Engineering Universitat Autónoma de Barcelona, Spain

[‡]Electrical & Computer Engineering Department, University of Houston, USA

[*]UNIK - University Graduate Center, University of Oslo, Norway

## Abstract

In this paper, we study multibeam satellite secure communication through physical (PHY) layer security techniques, i.e., joint power control and beamforming. By first assuming that the Channel State Information (CSI) is available and the beamforming weights are fixed, a novel secure satellite system design is investigated to minimize the transmit power with individual secrecy rate constraints. An iterative algorithm is proposed to obtain an optimized power allocation strategy. Moreover, sub-optimal beamforming weights are obtained by completely eliminating the co-channel interference and nulling the eavesdroppers' signal simultaneously. In order to obtain jointly optimized power allocation and beamforming strategy in some practical cases, e.g., with certain estimation errors of the CSI, we further evaluate the impact of the eavesdropper's CSI on the secure multibeam satellite system design. The convergence of the iterative algorithm is proven under justifiable assumptions. The performance is evaluated by taking into account the impact of the number of antenna elements, number of beams, individual secrecy rate requirement, and CSI. The proposed novel secure multibeam satellite system design can achieve optimized power allocation to ensure the minimum individual secrecy rate requirement. The results show that the joint beamforming scheme is more favorable than fixed beamforming scheme, especially in the cases of a larger number of satellite antenna elements and higher secrecy rate requirement. Finally, we compare the results under the current satellite air-interface in DVB-S2 and the results under Gaussian inputs.

# I. Introduction

The issues of privacy and security in satellite networks have taken on an increasing important role, especially in military applications. Currently, the secure satellite communication (SATCOM) is realized only through upper layer protocols (e.g., in [1], [2]). In this paper, we will investigate the multibeam satellite secure communication through physical (PHY) layer security techniques [3], [4], i.e., joint power control and beamforming schemes with individual secrecy rate constraints, which can be an alternative approach for satellite secure communication. Power limitation and co-channel interference are two challenges for multibeam satellite systems (e.g., in [5]–[8]). Hence, power control and beamforming could be two approaches for improving the system capacity by adjusting the beam pattern such that the overall transmitted power is minimized or the Signal-to-Interference plus Noise Ratio (SINR) is maximized.

In this paper, we consider the joint power and beamforming with individual secrecy rate constraints. An iterative algorithm is proposed for updating the transmission power in each iteration, such that a target secrecy rates are achieved for each beam with minimal power consumption. We first study the secure SATCOM system design through a power control problem with fixed beamforming. Next, the beamforming weights are achieved by co-channel interference cancelation and nulling the signal at the eavesdropper. Moreover, the impact of Channel State Information (CSI) of eavesdropper on secure SATCOM system design is studied.

In addition to security issues, the efficient resources management is also important for the SATCOM systems, e.g., bandwidth and power allocation. The authors n [9], [10] investigate the dynamic bandwidth allocation techniques for satellite systems. For the power control techniques in the satellite scenario, a power allocation policy is proposed in [5], which suggests to stabilize the system based on the amount of packets in the queue and the channel state, and a routing decision is made for the maximum total throughput. In [11], a tradeoff strategy is proposed between different objectives and system optimization. However, the co-channel interference is not taken into account and a convex optimization problem is solved. A joint power and carrier allocation problem is discussed in [6], however, only uplink is considered. In [7], [8], the authors focus on the capacity optimization in multibeam satellite system, and the duality of in frequency and time domain is studied. The optimization problem of power and carrier allocation has been addressed in terrestrial networks (e.g. [12], [13]). The authors in [13] propose an axiomatic-based interference model for SINR balancing problem with individual target SINR per user, but the conclusions are not directly extrapolable to a satellite scenario. To the best of our knowledge, the security issue is not discussed together with power control and beamforming in SATCOM systems. Beamforming is a sub-optimal strategy to reduce co-channel interference, but it has reduced complexity compared to Dirty-Paper Coding (DPC). In [14] and [15], transmit beamforming has been used to null the signal for each co-channel receiver. In [16], the authors studied the Zero-Forcing Beam-Forming (ZFBF) in the scenario of multiantenna broadcast where the weights are selected such as the multi-user co-channel interference is cancelled (zero-interference condition). Our work is different from the aforementioned literatures, since we introduce the physical (PHY) layer security for multibeam satellite systems and focus on the power control and co-channel interference management *jointly*.

Previous work (in [5]–[8], [11]–[15]) addresses the problems of power control by SINR balancing and beamforming separately, and without taking into account the secure communication issues. For security in SATCOM networks, there exits various works ( [1], [2]). However, most of it only focus on the upper layer security and realize through protocols, e.g., Authentication, Authorization, and Accounting Protocols (AAA), Transport Layer Security protocol (TLS), IP Security (IPSEC), Point-to-Point Tunnelling Protocol (PPTP), Internet Keying Exchange (IKE), and Internet Security Association and Key Management Protocol (ISAKMP) (in [2]). The PHY layer based security of wireless communication has been investigated since the contributions in [3], [4]. Recently, the application of PHY layer security in wireless communication is attracting more attention. E.g., in [17]–[19], the relay cooperating schemes are studied in order to maximize the achievable secrecy rate or minimize the transmit power. All the relays forward a weighted version of the decoded/amplified signal to the destination, thus, a maximized secrecy rate or minimized transmit power can be achieved by optimizing the weighting factor of each relay. The authors in [20] generalize the secure communication over the fading channels, the power allocation is derived to minimize the outage probability. Some recent work (in [21]–[28]) has been proposed to improve the performance, e.g., achievable secrecy rate, by taking advantage of multiple antenna systems. The authors in [21]–[25] investigate the PHY layer security by using Multiple-Input Multiple-Output (MIMO) systems. In [26], [27], the authors study the achievable rates in Gaussian Multiple-Input Single-Output (MISO) channels with secrecy constraints and conclude that the optimal solution can be achieved by beamforming in terms of the input covariance matrices. The Single-Input Multiple-Output (SIMO) case is studied in [28].

The main contributions of this paper are:

- We apply the PHY layer security in SATCOM scenarios, which is novel in the satellite networks. Since currently the security SATCOM is realized through upper layer protocols.

- We model the system as a MISO wiretap channel, which is different from the aforementioned

papers in various aspects. Existing MIMO/MISO models focus on the antenna-level for the terrestrial networks, while we focus on the beam-level for multibeam SATCOM systems. It means that, for a specific ground terminal, it corresponds to a specific beam on the satellite, the received signals by this terminal from other beams are considered as co-channel interference.

- The nature of the studied problem is different from the previous works. Existing works focused on the analysis of the achievable secrecy rate. Our aim is to characterize the secure SATCOM system through PHY layer design, i.e., power allocation and beamforming design under the individual secrecy rate constraints.

The main results of this paper can be summarized as:

- We prove that the proposed novel multibeam SATCOM system design can achieve the secure communication by jointly optimizing the power allocation and beamforming. As expected, in order to achieve the target secrecy rate, more power will be consumed in the cases of worse legitimate users' CSI and better eavesdropper's CSI.

- Two schemes, power control with fixed beamforming and with joint beamforming, are investigated and compared. We show that the joint beamforming scheme is more favorable than the fixed beamforming scheme, especially in the cases of a larger number of antenna elements and higher individual secrecy rate constraints.

- By comparing the results under the Gaussian inputs with the results under the current air-interface in DVB-S2, we come to the same conclusions.

The rest of this paper is organized as follows: In Section II, we model the multibeam downlink system to obtain a mathematical expression of the secrecy SINR and secrecy rate. The power control problem with fixed beamforming and iterative algorithm are studied in Section III. In Section IV, we propose and solve a joint power control and beamforming problem. The beamforming weight vector for each beam is obtained by joint ZFBF and eavesdropper nulling. The impact of the eavesdropper's CSI on the system design is presented in Section V. The performance of the algorithm and numerical results are presented in Section VI. In Section VII, we draw the conclusions.

We adopt the following notation: Bold uppercase letters denote matrices and bold lowercase letters denote vectors, $(\cdot)^*$, $(\cdot)^T$ and $(\cdot)^H$ denote conjugate, transpose, and conjugate transpose, respectively, $(\cdot)^{\dagger}$ denotes the Moore Penrose inverse, $\mathbb{E}\{\cdot\}$ denotes the expectation, $\text{var}\{x\}$ denotes the variance of $x$, $\text{diag}\{\mathbf{x}\}$ denotes a diagonal matrix with the elements of vector $\mathbf{x}$ along its main diagonal, $\mathbf{0}_{M \times N}$ denotes an all-zero matrix of size $M \times N$, $\|\mathbf{x}\|$ denotes the Euclidean norm of the vector $\mathbf{x}$, $\mathbf{I}_M$ is the identity matrix of size $M \times M$, $[\mathbf{X}]_{ij}$ denotes the $(i, j)$ entry of the matrix $\mathbf{X}$, $[\mathbf{x}]_j$ denotes the $j$th entry of the vector $\mathbf{x}$, and $\log(\cdot)$ denotes the base-2 logarithm.

## II. System Model

In the multibeam SATCOM system, we assume a security scenario (e.g., military application, as shown in Fig. 1), where only a few beams ($K$) are illuminated by coherently processing (e.g., beamforming) $M$ antenna elements. The $K$ illuminated beams serve $K$ decentralized legitimate users in the same frequency band. One eavesdropper, denoted $e$, is located outside/inside the satellite coverage. Both legitimate users and eavesdropper are assumed equipped with a single antenna. Therefore, for each of the specific user, the system can be seen as a MISO wiretap channel. It is different from the MISO model in [26], [27], because we focus on the beam-level and co-channel interference is taken into account. Our aim is to realize secure communication between the satellite and the legitimate users by transmit power control and beamforming. Next, we introduce the different sub-models.

### A. Channel Attenuation Amplitude Model

The attenuation due to the atmosphere depends on the frequency, the elevation angle, the altitude of the station, and the water vapor concentration [29]–[32]. As discussed in [33], the atmosphere attenuation (e.g., rain attenuation) is negligible at lower frequencies, e.g., less than 10 GHz, but has a strong impact on the performance at higher frequencies, e.g., Ka-band and above frequencies, which is the frequency band applied in current SATCOM systems [6]–[8]. Attenuation also depends on the distance that the electromagnetic wave propagates through space, i.e., path loss. We assume an instantaneous analysis with fixed channel transfer coefficients. The channel attenuation amplitude matrix $\mathbf{A} \in \mathbb{C}^{K \times K}$ is defined as

$$\mathbf{A} = \text{diag}\left\{\alpha_1, \alpha_2, \ldots, \alpha_K\right\}, \tag{1}$$

where $\alpha_i$ denotes the channel attenuation factor for legitimate user $i$ where $i = 1, 2, \ldots, K$. The channel attenuation factor for the eavesdropper is defined as $\alpha_e$.

### B. Antenna Model

We assume an Array Feed Reflector (AFR) antenna system [7], [8], [11], which is able to exploit the spatial characteristics of the propagation channel. Each beam is synthesized by adding array elements, hence, we can provide flexible power allocation by controlling the On-Board Processor (OBP). The array antenna system can achieve large performance gains, depending on the number of antenna elements and their relative position in space. However, these gains come at the cost of the increased hardware complexity. We suppose that the antenna gain matrix $\mathbf{G}$ of size $M \times K$ is given as

$$\mathbf{G} = \begin{bmatrix} g_{11} & g_{12} & \cdots & g_{1K} \\ g_{21} & g_{22} & \cdots & g_{2K} \\ \vdots & \vdots & \ddots & \vdots \\ g_{M1} & g_{M2} & \cdots & g_{MK} \end{bmatrix},$$

where $g_{ij}$ is the square root of the gain between the $i$th beam on-board antenna element and the $j$th legitimate user. The antenna gain between $M$ antenna elements and the eavesdropper is $\mathbf{g}_e = [g_{1e}, g_{2e}, \ldots, g_{Me}]^T$.

### C. Overall Channel Model

Let $\mathbf{H} = \mathbf{GA}$ be the overall channel matrix ($M \times K$) for the legitimate users, and let $\mathbf{h}_e = \alpha_e \mathbf{g}_e$ be the overall channel gain vector ($M \times 1$) between $M$ antenna elements and the eavesdropper. The overall channel between the satellite antenna elements and the legitimate users can be estimated accurately, e.g., by introducing a feedback channel. However, in practice, the channel condition between the satellite antenna elements and the eavesdropper is difficult to be estimated or is even totally unknown. In [34], the authors studied the possibility to estimate perfectly the CSI of the eavesdropper. However, it is only applicable in networks combining multicast and unicast transmissions, in which the terminals play dual roles as legitimate users for some signals and eavesdroppers for others.

In this paper, we do not focus on the CSI estimation, however, we study the system design of power control and beamforming with given CSI knowledge. Specifically, the two cases of complete eavesdropper CSI knowledge and imperfect eavesdropper CSI knowledge, are discussed. The first case is a common assumption in the PHY security literatures [19]–[21]. The attenuation factor can be estimated for the eavesdropper according to the method proposed in [29]–[32]. For the second case, which is more realistic in practical, we assume that only imperfect estimates of the eavesdropper's CSI are available.

### D. Received Signal Model

Let $s_k$ be the transmitted data symbol to User $k$. The amplitude of the signal transmitted to each user is normalized to one, i.e, $\mathbb{E}\{|s_k|^2\} = 1, \ \text{for} \ k = 1, 2, \ldots, K$. We denote by $P_k$ the allocated power to the $k$th beam and, hence, $\mathbf{p} = [P_1, P_2, \ldots, P_K]^T$ is the power allocation vector to all the beams. All signals are mapped onto the antenna array elements by the beamforming vectors $\mathbf{w}_k \in \mathbb{C}^{M \times 1}, \ \text{for} \ k = 1, 2, \ldots, K$.

Hence, the beamforming matrix $\mathbf{W} \in \mathbb{C}^{M \times K}$ is given by $\mathbf{W} = [\mathbf{w}_1, \mathbf{w}_2, \ldots, \mathbf{w}_K]$. Without loss of generality, we assume that $\|\mathbf{w}_k\| = 1$, for $k = 1, 2, \ldots, K$. Under this assumption, the transmitted power for each beam (e.g., beam $k$) is given by $P_k \mathbb{E}\{|s_k|^2\} = P_k$.

The block matrix model of the satellite broadcast scenario is shown in Fig. 2. The signals received by the $k$th user can be expressed as desired signal and interference as

$$y_k = \sqrt{P_k}\mathbf{h}_k^T\mathbf{w}_k s_k + \sum_{j \neq k} \sqrt{P_j}\mathbf{h}_k^T\mathbf{w}_j s_j + n_k, \tag{2}$$

where $\mathbf{h}_k$ (the $k$th column of $\mathbf{H}$) is the channel vector $(M \times 1)$ between $M$ antenna elements and the user in the $k$th ground cell. $n_k$ is signal-independent zero-mean complex circular Gaussian noise with variance $\sigma_k^2$ at beam $k$.

The signal received by the eavesdropper is given as

$$y_e = \sqrt{P_k}\mathbf{h}_e^T\mathbf{w}_k s_k + \sum_{j \neq k} \sqrt{P_j}\mathbf{h}_e^T\mathbf{w}_j s_j + n_e, \tag{3}$$

where the term $\sqrt{P_k}\mathbf{h}_e^T\mathbf{w}_k s_k$ is the desired signal if the eavesdropper intend to wiretap the $k$th user. $\mathbf{h}_e^T\mathbf{w}_j$ denotes the channel gain between the eavesdropper and the $j$th antenna element, and $n_e$ is a zero-mean complex circular Gaussian noise at the eavesdropper. We assume that the noise level at the legitimate user (e.g., user $k$) is equal to that of at the eavesdropper, i.e., $\text{var}\{n_k\} = \text{var}\{n_e\} = \sigma^2, \forall k$ . This is a reasonable assumption since the sensitivity of all the terminals is often similar.

### E. Signal-to-Interference Plus Noise Ratio

Let $\mathbf{R}_k \triangleq (\mathbf{h}_k\mathbf{h}_k^H)^T$, for $k = 1, 2, \ldots, K$, and $\mathbf{R}_e \triangleq (\mathbf{h}_e\mathbf{h}_e^H)^T$. According to the formulation of the received signal in (2) and (3), we can derive the SINR of the legitimate user $k$ as

$$\Gamma_k = \frac{P_k\mathbf{w}_k^H\mathbf{R}_k\mathbf{w}_k}{\sigma^2 + \sum_{j \neq k} P_j\mathbf{w}_j^H\mathbf{R}_k\mathbf{w}_j}, \tag{4}$$

and the SINR of the eavesdropper, which intend to wiretap the signal transmitted to user $k$ as

$$\Gamma_{ek} = \frac{P_k\mathbf{w}_k^H\mathbf{R}_e\mathbf{w}_k}{\sigma^2 + \sum_{j \neq k} P_j\mathbf{w}_j^H\mathbf{R}_e\mathbf{w}_j}. \tag{5}$$

### F. Secrecy Rate Model

As we have indicated in the introduction, there have been several precedents that investigate the PHY layer security of the MIMO wiretap channel, but they only focus on the terrestrial networks. Certainly, these results also cover the special case of the MISO channel. For the case of one eavesdropper, an achievable secrecy rate for a specific user (e.g., for the $k$th user) is given as [19, Eq. (10)]

$$R_s^k = \max\{R_k - R_{ek}\}, \tag{6}$$

where the achievable of the maximum was shown in [25], [27] with Gaussian inputs, $R_k$ is the achievable rate of the link between the satellite and the $k$th user, and $R_{ek}$ is the achievable rate of the link between the satellite and the eavesdropper. Note that the secrecy rate in (6) is achievable unless the maximum value is negative, in which case, the achieved secrecy rate is zero [4]. In this paper, we focus on the practical scenario in which the secrecy rate is positive.

In [26], [27], the authors discuss how to maximize the difference by adaptively adjust the power allocation. Conversely, we restrict ourselves to the difference between $R_k$ and $R_{ek}$. Our aim is to characterize the best power allocation scheme over multibeam SATCOM systems subject to the individual secrecy rate constraints, i.e. the difference $R_k - R_{ek}$ for each user.

By assuming Gaussian inputs, the difference between $R_k$ and $R_{ek}$ can be written as

$$R_k - R_{ek} = \log\left(1 + \Gamma_k\right) - \log\left(1 + \Gamma_{ek}\right) = \log\frac{1 + \Gamma_k}{1 + \Gamma_{ek}} = \log\left(1 + \frac{\Gamma_k - \Gamma_{ek}}{1 + \Gamma_{ek}}\right) = \log\left(1 + \Gamma_s^k\right), \tag{7}$$

where $\Gamma_s^k$ is defined as the secrecy SINR, which is the updated SINR after introducing the eavesdropping, and it is given by

$$\Gamma_s^k \triangleq \frac{\Gamma_k - \Gamma_{ek}}{1 + \Gamma_{ek}}. \tag{8}$$

From (4) and (5), we notice that $\Gamma_s^k$ is a function of two parameters, i.e., the beamforming matrix $\mathbf{W}$ and the power vector $\mathbf{p}$. In the next sections, we will discuss how to minimize the overall power consumption (sum of the elements inside $\mathbf{p}$) under the SINR constraint per beam by taking into account both fixed and optimized beamforming matrix. From (7), we can see that the optimization problem with a secrecy SINR constraint is equivalent to the secrecy rate constraint. If we consider that the secrecy rate required by the $k$th user is $\hat{R}_s^k$, the secrecy SINR requirement can be derived as $\gamma_k = 2^{\hat{R}_s^k} - 1$. Therefore, in the following section, we focus on the power control problem with a secrecy SINR constraint per user.

## III. Power Control Problem with Fixed Beamforming

In this section, we assume that the beamforming matrix $\widetilde{\mathbf{W}} = [\widetilde{\mathbf{w}}_1, \widetilde{\mathbf{w}}_2, \ldots, \widetilde{\mathbf{w}}_K]$ is optimized, with $\|\widetilde{\mathbf{w}}_k\| = 1$, for $k = 1, 2, \ldots, K$. We focus on the secure SATCOM system design through power allocation with individual SINR constraints.

A more general solution based on [35] is proposed to solve the power control problem. By doing the multibeam satellite power control, the overall transmit power of each beam is optimized, so that the received secrecy rate of each user satisfies $R_s^k \geq \hat{R}_s^k$, for $k = 1, 2, \ldots, K$, i.e., the secrecy SINR has $\Gamma_s^k \geq \gamma_k$ for $k = 1, 2, \ldots, K$, (where $\gamma_k$ is the predefined targeted SINR threshold in order to realize the required secrecy rate), while the overall transmitted power used by all beams is minimized. Hence, the power control problem can be defined as

$$\min_{\mathbf{p}} \ \sum_k P_k, \tag{9}$$
$$\text{subject to } \Gamma_s^k(\widetilde{\mathbf{W}}, \mathbf{p}) \geq \gamma_k, \ \ k = 1, 2, \ldots, K.$$

The minimum power is achieved when the SINR is equal to the target value, i.e., $\Gamma_s^k = \gamma_k$ for $k = 1, 2, \ldots, K$. The problem in (9) is a Nondeterministic Polynomial (NP) hard problem [36, Chapter 5, pp. 109]. Therefore, an iteration algorithm is proposed to find a solution. Many iteration algorithms (e.g., in [37]–[39]) have been proposed in order to decrease the complexity. However, the algorithm in this paper is different, since the eavesdropper is present.

We first construct the complete iteration expression as $\mathbf{I}(\mathbf{p}^n)$, which is a power-update equation. $\mathbf{p}^n = [P_1^n, P_2^n, \ldots, P_K^n]^T$ is the power vector for all the $K$ beams at the $n$th iteration step. Then, for each beam (e.g., beam $k$), the interference function $I_k(\mathbf{p})$ can be derived. The power allocated to each beam can be iteratively updated until converge with the individual secrecy SINR constraints. The algorithm steps at the $(n+1)$th iteration are as follows:

*Iteration Algorithm:*

$$\mathbf{p}^{n+1} = \mathbf{I}(\mathbf{p}^n) \tag{10}$$

The power-update for the $k$th beam at the $(n+1)$th iteration is

$$P_k^{n+1} = \frac{\gamma_k}{\mu_k^n - (1 + \gamma_k)\mu_{ek}^n} \triangleq I_k(\mathbf{p}^n), \tag{11}$$

where $\mu_k^n$ and $\mu_{ek}^n$ are defined as

$$\mu_k^n = \frac{\Gamma_k^n}{P_k^n} = \frac{\Theta_{kk}}{\sigma^2 + \sum_{j \neq k} P_j^n \Theta_{kj}}, \tag{12}$$

and

$$\mu_{ek}^n = \frac{\Gamma_{ek}^n}{P_k^n} = \frac{\Theta_{ek}}{\sigma^2 + \sum_{j \neq k} P_j^n \Theta_{ej}}, \tag{13}$$

respectively, where $\Gamma_k^n$ and $\Gamma_{ek}^n$ are the updated SINR of the legitimate user $k$ and the eavesdropper at the $n$th iteration step, $\Theta_{kk} = \widetilde{\mathbf{w}}_k^H \mathbf{R}_k \widetilde{\mathbf{w}}_k$, $\Theta_{kj} = \widetilde{\mathbf{w}}_j^H \mathbf{R}_k \widetilde{\mathbf{w}}_j$, $\Theta_{ek} = \widetilde{\mathbf{w}}_k^H \mathbf{R}_e \widetilde{\mathbf{w}}_k$, and $\Theta_{ej} = \widetilde{\mathbf{w}}_j^H \mathbf{R}_e \widetilde{\mathbf{w}}_j$.

The computation of $\Theta_{kk}$, $\Theta_{kj}$, $\Theta_{ek}$, and $\Theta_{ej}$ dominates the computational complexity of the algorithm. Since $\widetilde{\mathbf{w}}_k$ is a $M \times 1$ vector, $\mathbf{R}_k$ and $\mathbf{R}_e$ are $M \times M$ matrices, thus, the expressions in (12) and (13) require a computational complexity of $O(M^4)$ for updating the allocated power per user. Thus, the computational complexity is quite high for the cases of large number of beams. However, in this paper, we assume a security scenario (e.g., military application), where only a few beams are implemented, hence, the number of antenna element $M$ is quite low (e.g., max. 20 as we assumed), and the proposed algorithm computational complexity is reasonable. In addition, since the satellite channel is relatively stable, the computations needed are less frequently. Moreover, as we have noted in the antenna model section, although the array antenna system can achieve large performance gains with large number of antenna elements $M$, these gains come at the cost of the increased hardware complexity and computational complexity.

In [35], [40], the authors have proved that if the interference function is *standard*, the algorithm will achieve the optimal solution if there exists at least one feasible solution. The interference function $I_k(\mathbf{p})$ is *standard* if for all $\mathbf{p} \geq 0$ the following three properties are satisfied [35], [40]:

- Positivity: $I_k(\mathbf{p}) > 0$.

- Monotonicity: If $\mathbf{p} \geq \mathbf{p}'$, then $I_k(\mathbf{p}) \geq I_k(\mathbf{p}')$ or $I_k(\mathbf{p}) \leq I_k(\mathbf{p}')$.[1]

- Scalability: For all $\rho > 1$, $\rho I_k(\mathbf{p}) > I_k(\rho \mathbf{p})$.

For the proposed interference function (11), we obtain the following theorem:

*Theorem 1:* The interference function $I_k(\mathbf{p}^n)$ in (11) is a *standard* function under the following three conditions:

- Condition 1: $b_k > c_k$.

- Condition 2: $b_k \tilde{\mathbf{h}}_k > c_k \tilde{\mathbf{h}}_e$, $b_k \tilde{\mathbf{h}}_e > c_k \tilde{\mathbf{h}}_k$, and $b_k \tilde{\mathbf{h}}_k \tilde{\mathbf{h}}_k^T > c_k \tilde{\mathbf{h}}_e \tilde{\mathbf{h}}_e^T$, $b_k \tilde{\mathbf{h}}_e \tilde{\mathbf{h}}_e^T > c_k \tilde{\mathbf{h}}_k \tilde{\mathbf{h}}_k^T$, $\forall k$.[2]

- Condition 3: $\sqrt{b_k [\tilde{\mathbf{h}}_{kk}]_j} \tilde{\mathbf{h}}_e > \sqrt{c_k [\tilde{\mathbf{h}}_e]_j} \tilde{\mathbf{h}}_k$, $\forall k, j \neq k$.

Where $b_k = \Theta_{kk}$, $c_k = (1 + \gamma_k)\Theta_{ek}$, and $\tilde{\mathbf{h}}_k$ denotes the channel gain vector ($K \times 1$) of the interference

---

[1] The inequality between two vectors, e.g., $\mathbf{x} \geq \mathbf{y}$, means that $x_i \geq y_i$ for $i = 1, \ldots, K$, where $\mathbf{x} = [x_1, x_2, \ldots, x_K]$, $\mathbf{y} = [y_1, y_2, \ldots, y_K]$.

[2] The inequality between two matrices, e.g., $\mathbf{X} \geq \mathbf{Y}$, means that $[\mathbf{X}]_{ij} > [\mathbf{Y}]_{ij}$, $\forall i, j$.

contribution to the desired user, defined as

$$[\tilde{\mathbf{h}}_k]_j = \begin{cases} \Theta_{kj}, & \text{if } j \neq k, \\ 0, & \text{otherwise.} \end{cases}$$

$\tilde{\mathbf{h}}_e$ denotes the channel gain vector $(K \times 1)$ of the interference contribution to the eavesdropper, defined as

$$[\tilde{\mathbf{h}}_e]_j = \begin{cases} \Theta_{ej}, & \text{if } j \neq k, \\ 0, & \text{otherwise.} \end{cases}$$

The proof Theorem 1 is presented in Appendix A.

In a practical scenario, the overall channel gain of the link "satellite - desired user" is much larger than that of the link "satellite - co-channel users", i.e., $\Theta_{kk} \gg \Theta_{kj}$ for $\forall j \neq k$, the overall channel gain of the link "satellite - desired user" is larger than that of the link "satellite - eavesdropper", i.e., $\Theta_{kk} \gg \Theta_{ej}$ for $\forall j$. The magnitudes of $\Theta_{kj}$ and $\Theta_{ej}$ are roughly equal. Therefore, with the lower secrecy SINR request $\gamma_k$, the above three conditions are indeed satisfied. In the case of very high SINR requirement, we can introduce optimization of the satellite antenna beamformer in order to decrease or eliminate the co-channel interference and the eavesdropper interference, and thereby the above conditions can still be satisfied.

## IV. Joint Power Control and Beamforming

The level of co-channel interference and wiretapped signal for each user depend both on the gain between interfering transmitters and user, as well as on the level of transmitter powers, i.e., the optimal beamforming vector may vary for different power allocation policy. Hence, in this section, we first obtain a sub-optimal beamforming weight vector by completely eliminating the co-channel interference and nulling the eavesdroppers' signal simultaneously. Then, the power solution can be optimized when the secrecy rate is equal to the target value.

In the joint power control and beamforming problem, the objective is to find the optimal weight matrix $\mathbf{W}$ and power allocation vector $\mathbf{p}$ such that the secrecy SINR threshold is achieved by all the users, while minimize the transmission power. The problem can be formulated as

$$\min_{\mathbf{W},\mathbf{p}} \sum_k P_k, \tag{14}$$

subject to $\Gamma_s^k(\mathbf{W}, \mathbf{p}) \geq \gamma_k, \quad k = 1, 2, \ldots, K.$

This problem can be solved in two steps: Firstly, we obtain the beamforming weight matrix $\mathbf{W}$ by joint ZFBF and eavesdropper signal nulling, in which all the co-channel signal and eavesdropper signal are completely eliminated. Secondly, the optimized power allocation solution can be obtained by solving $\Gamma_s^k = \gamma_k$ for $k = 1, 2, \ldots, K$, under the beamforming weights obtained in the first step.

### A. Joint Zero-Forcing Beamforming and Eavesdropper Nulling

In order to completely eliminate the co-channel interference and null the signals at the eavesdropper, we assume that $M > K$. Note that in the case of $M \leq K$, we cannot completely eliminate the interference from the co-channel users and nulling the signals at the eavesdropper; appropriate system design for the case of $M \leq K$ is an interesting future research direction.

By ZFBF (in [41], [42]), the weights are selected such as the co-channel interference is canceled (zero-interference condition), i.e., for the desired user $k$, $\mathbf{h}_k^T \mathbf{w}_j = 0$ for $j \neq k$. Similarly, the eavesdropping interference can also be completely nulled by beamforming (e.g., in [17]–[19]), i.e., for the desired user $k$, $\mathbf{h}_e^T \mathbf{w}_k = 0$ for $k = 1, 2, \ldots, K$.

Hence, the secrecy SINR can be reformulated from (8) as

$$\Gamma_s^k(\mathbf{W}, \mathbf{p}) = \frac{P_k \mathbf{w}_k^H \mathbf{R}_k \mathbf{w}_k}{\sigma^2} = \frac{P_k |\mathbf{h}_k^T \mathbf{w}_k|^2}{\sigma^2}. \tag{15}$$

Therefore, in order to minimize the transmitted power $P_k$, for $k = 1, 2, \ldots, K$, under the secrecy SINR constraints $\gamma_k$, we have to maximize the gain between the satellite antenna and the $k$th user, i.e., $\max |\mathbf{h}_k^T \mathbf{w}_k|^2$, for $k = 1, 2, \ldots, K$. It means that we have to solve $K$ maximize problems jointly. The $k$th optimization problem can be formulated as

$$\max_{\mathbf{w}_k} \ |\mathbf{h}_k^T \mathbf{w}_k|^2, \tag{16}$$

$$\text{subject to} \begin{cases} \mathbf{h}_k^T \mathbf{w}_j = 0, \ \text{for} \ j \neq k, \\ \mathbf{h}_e^T \mathbf{w}_k = 0, \\ \mathbf{w}_k^H \mathbf{w}_k = 1. \end{cases}$$

Note that the overall optimization problem is composed of $K$ optimization problems as expressed in (16) (for $k = 1, 2, \ldots, K$). In an equivalent way, we re-formulate the $K$ *jointly* maximize problems as $K$ *independent* maximization problem, e.g., the problem to solve the $k$th beamforming weight vector can

be formulated as

$$\max_{\mathbf{w}_k} \ |\mathbf{h}_k^T \mathbf{w}_k|^2, \tag{17}$$

$$\text{subject to} \begin{cases} \mathbf{H}_{ek}^T \mathbf{w}_k = \mathbf{0}_{K \times 1}, \\ \mathbf{w}_k^H \mathbf{w}_k = 1, \end{cases}$$

where $\mathbf{H}_{ek}$ is defined as

$$[\mathbf{H}_{ek}]_{ij} \triangleq \begin{cases} [\mathbf{H}]_{ij}, & \text{if } j \neq k, \\ [\mathbf{h}_e]_i, & \text{if } j = k. \end{cases} \tag{18}$$

The solution of the beamforming problem in (17) is given by [19, Eq. (23)] as

$$\mathbf{w}_k = \frac{(\mathbf{I}_M - \mathbf{F}_e) \, \mathbf{h}_k^*}{\| (\mathbf{I}_M - \mathbf{F}_e) \, \mathbf{h}_k^* \|}, \ \text{ for } \ k = 1, 2, \ldots, K, \tag{19}$$

where

$$\mathbf{F}_e = (\mathbf{H}_{ek})^\dagger \mathbf{H}_{ek},$$

where $(\mathbf{H}_{ek})^\dagger = (\mathbf{H}_{ek})^H \left( \mathbf{H}_{ek} (\mathbf{H}_{ek})^H \right)^{-1}$ is the Moore Penrose inverse of $\mathbf{H}_{ek}$ (in [43]).

As discussed in Section III, the minimum power is achieved when the SINR is equal to the target value, i.e., $\Gamma_s^k = \gamma_k$ for $k = 1, 2, \ldots, K$. Therefore, we can obtain the solution from (15) as

$$P_k = \frac{\gamma_k \sigma^2}{|\mathbf{h}_k^T \mathbf{w}_k|^2}, \ \text{ for } \ k = 1, 2, \ldots, K, \tag{20}$$

where $\mathbf{w}_k$ is the solution of the beamforming weight vector for the $k$th beam.

## V. Impact on CSI of Eavesdropper

The channels between the satellite and the desired users can be estimated accurately, since they are legitimate channels. However, in practice, the channels between the satellite and the eavesdropper can only be estimated, and the estimation contains errors. In the following two subsections, we will investigate the system design with unknown or imperfect CSI of the eavesdropper.

### A. Unknown Eavesdropper CSI

In this case, we assume that the entries of $\mathbf{h}_e$ are random variables, and $\widehat{\mathbf{R}}_e = \mathbb{E} \left\{ (\hat{\mathbf{h}}_e \hat{\mathbf{h}}_e^H)^T \right\}$ is known a priori. Therefore, in order to minimize the power consumption subject to given target secrecy SINR, we can use a sub-optimal way to cancel the co-channel interference, i.e., ZFBF.

We can formulate the $k$th beamforming weight vector optimization problem as

$$\max_{\mathbf{w}_k} \ |\mathbf{h}_k^T \mathbf{w}_k|^2, \tag{21}$$

$$\text{subject to} \begin{cases} \mathbf{h}_k^T \mathbf{w}_j = 0, \ \text{ for } \ j \neq k, \\ \mathbf{w}_k^H \mathbf{w}_k = 1. \end{cases}$$

This problem is similar to the problem formulated in (16), thus, we obtain the solution as

$$\mathbf{w}_k = \frac{(\mathbf{I}_M - \mathbf{F}) \, \mathbf{h}_k^*}{\| \, (\mathbf{I}_M - \mathbf{F}) \, \mathbf{h}_k^* \|}, \ \text{ for } \ k = 1, 2, \ldots, K, \tag{22}$$

where

$$\mathbf{F} = (\mathbf{H}_k)^{\dagger} \mathbf{H}_k,$$

where $(\mathbf{H}_k)^{\dagger} = (\mathbf{H}_k)^H \left( \mathbf{H}_k (\mathbf{H}_k)^H \right)^{-1}$, and $\mathbf{H}_k$ is the co-channel contribution matrix $M \times (K-1)$ defined as

$$\mathbf{H}_k \triangleq [\mathbf{h}_1, \ldots, \mathbf{h}_{k-1}, \mathbf{h}_{k+1}, \ldots, \mathbf{h}_K], \tag{23}$$

where $\mathbf{h}_j \ (j \neq k)$ is the $j$th column of the channel matrix $\mathbf{H}$.

After obtain the beamforming vector for each beam, the power allocation solution can also be obtained by the iteration algorithm in (11), i.e.,

$$P_k^{n+1} = \frac{\gamma_k}{\mu_k^n - (1 + \gamma_k) \mu_{ek}^n}, \tag{24}$$

where $\mu_k^n$ and $\mu_{ek}^n$ are re-defined in *Theorem* 2.

*Theorem 2:* The interference function in (24) is a *standard* function under the condition: $b_k > c_k$, where $b_k = \mathbf{w}_k^H \mathbf{R}_k \mathbf{w}_k$, $c_k = (1 + \gamma_k) \mathbf{w}_k^H \widehat{\mathbf{R}}_e \mathbf{w}_k$. $\mu_k^n$ and $\mu_{ek}^n$ are defined as

$$\mu_k^n = \frac{\mathbf{w}_k^H \mathbf{R}_k \mathbf{w}_k}{\sigma^2}, \tag{25}$$

and

$$\mu_{ek}^n = \frac{\mathbf{w}_k^H \widehat{\mathbf{R}}_e \mathbf{w}_k}{\sum_{j \neq k} P_j^n \mathbf{w}_j^H \widehat{\mathbf{R}}_e \mathbf{w}_j + \sigma^2}, \tag{26}$$

respectively.

See Appendix B for the proof of Theorem 2.

## B. Imperfect Eavesdropper CSI

The perfect channel gain $\mathbf{h}_e \in \mathbb{C}^{M \times 1}$ between the satellite antenna elements and eavesdropper is modeled as

$$\mathbf{h}_e = \hat{\mathbf{h}}_e + \boldsymbol{\Delta}_e, \tag{27}$$

where $\hat{\mathbf{h}}_e \in \mathbb{C}^{M \times 1}$ is the imperfect eavesdropper channel estimation, and $\boldsymbol{\Delta}_e \in \mathbb{C}^{M \times 1}$ corresponds to the channel estimation error. We assume that the entries of $\boldsymbol{\Delta}_e$ are random variables, which is independent of $\hat{\mathbf{h}}_e$, and $\mathbf{R}_{\Delta_e} \triangleq \mathbb{E} \left\{ (\boldsymbol{\Delta}_e \boldsymbol{\Delta}_e^H)^T \right\}$ is known a priori. Thus,

$$\mathbf{R}_e = \mathbb{E} \left\{ (\mathbf{h}_e \mathbf{h}_e^H)^T \right\} = \hat{\mathbf{R}}_e + \mathbf{R}_{\Delta_e}, \tag{28}$$

where $\hat{\mathbf{R}}_e = (\hat{\mathbf{h}}_e \hat{\mathbf{h}}_e^H)^T$.

By joint ZFBF and nulling the eavesdropper's signal, we obtain the beamforming vector, e.g., for the $k$th beam, as expressed in function (19). However, $\mathbf{H}_{ek}$ is replaced with $\hat{\mathbf{H}}_{ek}$, which is defined as

$$[\hat{\mathbf{H}}_{ek}]_{ij} = \begin{cases} [\mathbf{H}]_{ij}, & \text{if } j \neq k, \\ [\hat{\mathbf{h}}_e]_i, & \text{if } j = k. \end{cases} \tag{29}$$

We can solve the power control problem with the iteration algorithm in function (24), then $\mu_{ek}^n$ is re-defined as

$$\mu_{ek}^n = \frac{\mathbf{w}_k^H \mathbf{R}_{\Delta_e} \mathbf{w}_k}{\displaystyle\sum_{j \neq k} P_j^n \mathbf{w}_j^H \mathbf{R}_{\Delta_e} \mathbf{w}_j + \sigma^2}. \tag{30}$$

As expressed in *Theorem* 2, the interference function in (24) is *standard* with $\mu_k^n$ and $\mu_{ek}^n$ given in (25) and (30), respectively.

## VI. SIMULATION RESULTS ANALYSIS

In order to evaluate the performance of the proposed system designs, we perform Monte Carlo simulations consisting of 1000 independent trials to obtain the average results. We define the SATCOM system payload parameters the same as in [7] and assume that the noise power $\sigma^2$ is -10 dBm. For simplicity, the secrecy SINR request for all the beams is assumed to be equal, i.e., $\gamma_k = \gamma_0$ for $k = 1, 2, \ldots, K$. The channel for each link is modeled as a product of an attenuation factor and a random phase. For example, the channel between the legitimate user $k$ and the antenna element $m$ is defined as $h_{mk} = \alpha_k e^{j\varsigma}$, and the channel between the antenna elements and the eavesdropper is $h_{em} = \alpha_e e^{j\varsigma}$, where $\varsigma$ is a random phase uniformly distributed within $[0, 2\pi)$, and it is independent of $m$ and $k$.

We first fix the number of antenna elements to $M = 8$, the number of beams to $K = 5$, the channel attenuation factor $\alpha_k = \alpha_e = 0.8$ for $k = 1, 2, \ldots, K$ to investigate the convergence of the iteration algorithm. In Fig. 3, the curves show the total power consumption at each iteration step for different target secrecy SINR. The results show that the algorithm converge. Notice from the figure that the black curve with higher target SINR ($\gamma_0 = 8$ dB) converges slower than that of the red curve with lower target SINR ($\gamma_0 = 6$ dB), since more power is needed to achieve higher SINR requirements.

Fig. 4 illustrates satellite transmit power versus the number of antenna elements $M$. The fixed beamforming vector (e.g., for beam $k$) is assumed as $\|\tilde{\mathbf{w}}_k\| = 1$. The curves show that the transmitted power in the scheme of fixed beamforming is almost independent of the number of antenna elements, and the transmitted power in the scheme of joint beamforming decreases as the number of antenna elements increases. From the optimization point of view, the satellite transmitted power can be saved by increasing the number of antenna elements. However, from the satellite payload designers' point of view, the complexity and the weight of the satellite will increase as the number of antenna elements increases. Therefore, the optimal number of antenna elements should be balanced by taking into account all these views.

In Fig. 5, we evaluate the transmitted power according to different number of beams on the satellite. We fix the number of antenna elements at $M = 15$ and increase the number of beams $K$ from 2 to 12. All other parameters are the same in Fig. 4. As expected, the power consumption increases as the number of beams and secrecy request increase for both schemes. Especially, the transmitted power increases very quickly in the case of a large number of beams. In Fig. 6, we simulate the power allocation according to the channel attenuation amplitude of the eavesdropper, the horizontal axis in the figure indicates the channel attenuation amplitude degradation in dB, e.g., 0 dB means the clear sky scenario. From the figure, we see that the joint beamforming scheme is almost independent of the eavesdropper's channel condition, which means that the satellite can adapt the channel degradation by optimizing the beamformer design. For the fixed beamforming scheme, the transmitted power will decrease as the eavesdropper's channel condition deteriorates.

The performance of the transmit power as a function of the secrecy SINR request is shown in Fig. 7. For simplicity, we assume that the channel attenuation amplitudes for all the users are the same, and the channel attenuation amplitude of the eavesdropper is assumed as $\alpha_e = 1$, clear sky. All other parameters are the same as previous figures. For both fixed beamforming and joint beamforming schemes, the curves in Fig. 7 show that, as the channel condition deteriorates, more power will be consumed in order to compensate the signal attenuation. We can also conclude from this figure that the joint beamforming scheme is more favorable than fixed beamforming scheme in the case of a higher secrecy SINR request, since the power allocation is more sensitive to the higher secrecy SINR request (e.g., when $\gamma_0 > 6$ dB).

The performance of a single legitimate user (e.g., User 1) is evaluated in Fig. 8. We assume that the channel attenuation amplitude of User 1 ($\alpha_1$) is changed from 1 (i.e., clear sky) to 0.2, and all other parameters are the same in Fig. 3. As expected, the power allocated to Beam 1 will increase as the channel condition of User 1 deteriorates, especially in the case of a bad channel condition. In Fig. 9, we compare the power allocation with and without the available of the eavesdropper's CSI. The value of the parameters is the same in Fig. 7. Under the given total power limitation (e.g., 100 Watts), the achieved secrecy SINR per user with known eavesdropper's CSI performs about 2 dB better than the case of no CSI available. In addition, this gap increases as the available total power increases.

In Fig. 10, we compare the results with Gaussian inputs and with the current air-interface in DVB-S2. The value of the parameters is assumed to be the same as in Fig. 7. For the case of the joint beamforming scheme, the sum of power consumption increases as the spectral efficiency requirement increases for both Gaussian inputs and DVB-S2 cases. The power consumption of the DVB-S2 case is always larger than the Gaussian inputs case, and the gap between them tends to decrease as the spectral efficiency increases.

Table I shows the maximum number of users for different system designs. We assume that $P_{\text{tot}} = 10$Watt, $\gamma_0 = 6$dB, and $M = 20$. The first row indicates the maximum capacity of the system design for a fixed power allocation and a fixed beamforming system design, which is the baseline reference system design. We can notice that the capacity of the system design only with the flexibility in power allocation is around two times better than the reference one, and the capacity of the joint power control and beamforming system design is five times better than the reference one. In addition, the table also shows that the capacity of the joint power control and beamforming system design is not sensitive to the eavesdropper's channel condition.

## VII. Conclusions

By PHY layer techniques, we realize secure communication of multibeam SATCOM systems while minimizing the overall transmitted power. The power control problems is developed in different cases and an iterative algorithm is proposed to solve the problems. Specifically, we first assume that the beamforming weights are fixed, and propose a novel secure SATCOM system design that minimizes the satellite transmit power with individual secrecy rate constraints. A joint power control and beamforming problem is investigated to realize secure communication. The beamforming weight vector is solved by completely eliminating the co-channel interference and nulling the eavesdroppers' signal simultaneously.

Furthermore, the impact of channel condition of eavesdropper on the secure system design is studied. After the numerical simulation, we conclude that the proposed multibeam SATCOM system design can realize the secure communication by joint power control and beamforming. In order to achieve the target individual secrecy rate per user, more power will be consumed in the cases of worse legitimate users' CSI and better eavesdropper's CSI. The results also show that the joint power and beamforming scheme is more favorable than the fixed beamforming scheme in the cases of larger number of antenna elements and higher secrecy SINR request. Under a given overall power limitation (e.g., 100 Watts), the maximum secrecy SINR achieved per user with known eavesdropper's CSI preforms 2 dB better than the case without CSI available. By comparing the results with Gaussian inputs and with the current air-interface in DVB-S2, we come to the same conclusions.

## Appendix A

## Proof of Theorem 1

### A. Proof of Positivity

The interference function $I_k(\mathbf{p})$ in (11) can be rewritten as

$$I_k(\mathbf{p}) = \frac{a_k}{\frac{b_k}{\sigma^2 + \mathbf{p}^T\tilde{\mathbf{h}}_k} - \frac{c_k}{\sigma^2 + \mathbf{p}^T\tilde{\mathbf{h}}_e}} = \frac{a_k}{f(\mathbf{p})}, \tag{31}$$

where $b_k$ and $c_k$ are defined in Section III, and $a_k = \gamma_k > 0$. $f(\mathbf{p})$ is defined as

$$f(\mathbf{p}) = \frac{b_k}{\sigma^2 + \mathbf{p}^T\tilde{\mathbf{h}}_k} - \frac{c_k}{\sigma^2 + \mathbf{p}^T\tilde{\mathbf{h}}_e} = \frac{\sigma^2\left(b_k - c_k\right) + \left(b_k\mathbf{p}^T\tilde{\mathbf{h}}_e - c_k\mathbf{p}^T\tilde{\mathbf{h}}_k\right)}{\left(\sigma^2 + \mathbf{p}^T\tilde{\mathbf{h}}_e\right)\left(\sigma^2 + \mathbf{p}^T\tilde{\mathbf{h}}_k\right)}. \tag{32}$$

Under the assumed conditions, we obtain $b_k - c_k > 0$ and $b_k\mathbf{p}^T\tilde{\mathbf{h}}_e - c_k\mathbf{p}^T\tilde{\mathbf{h}}_k > 0$. Therefore, $f(\mathbf{p}) > 0$, and the positivity is proved.

### B. Proof of Monotonicity

A preference operator or *equivalent relation* "⇔" is defined for indicating that two expressions are equivalent. E.g., "$I_k(\mathbf{p})$ monotonically increasing" ⇔ "$f(\mathbf{p})$ monotonically decreasing", where $f(\mathbf{p})$ is defined in (32).

Let $\boldsymbol{\varphi}(\mathbf{p})$ be defined as $\boldsymbol{\varphi}(\mathbf{p}) = \frac{\partial f(\mathbf{p})}{\partial \mathbf{p}}$ and, hence, "$f(\mathbf{p})$ monotonically decreasing" ⇔ "$\boldsymbol{\varphi}(\mathbf{p}) < \mathbf{0}$, if $\mathbf{p} > \mathbf{0}$". $\boldsymbol{\varphi}(\mathbf{p})$ can be formulated as

$$\boldsymbol{\varphi}(\mathbf{p}) = \frac{\partial f(\mathbf{p})}{\partial \mathbf{p}} = \frac{c_k\tilde{\mathbf{h}}_e}{\left(\sigma^2 + \mathbf{p}^T\tilde{\mathbf{h}}_e\right)^2} - \frac{b_k\tilde{\mathbf{h}}_k}{\left(\sigma^2 + \mathbf{p}^T\tilde{\mathbf{h}}_k\right)^2} = \frac{\boldsymbol{\psi}(\mathbf{p})}{\left(\sigma^2 + \mathbf{p}^T\tilde{\mathbf{h}}_k\right)^2\left(\sigma^2 + \mathbf{p}^T\tilde{\mathbf{h}}_e\right)^2}, \tag{33}$$

where

$$\psi(\mathbf{p}) = c_k \tilde{\mathbf{h}}_e \Big( \sigma^2 + \mathbf{p}^T \tilde{\mathbf{h}}_k \Big)^2 - b_k \tilde{\mathbf{h}}_k \Big( \sigma^2 + \mathbf{p}^T \tilde{\mathbf{h}}_e \Big)^2. \tag{34}$$

Thus, "$\boldsymbol{\varphi}(\mathbf{p}) < \mathbf{0}$, if $\mathbf{p} > \mathbf{0}$" $\Leftrightarrow$ "$\boldsymbol{\psi}(\mathbf{p}) < \mathbf{0}$, if $\mathbf{p} > \mathbf{0}$". For the $j$th element of $\boldsymbol{\psi}(\mathbf{p})$, i.e., $\boldsymbol{\psi}_j(\mathbf{p})$, it can be presented as

$$\boldsymbol{\psi}_j(\mathbf{p}) = 2c_k [\tilde{\mathbf{h}}_e]_j \Big( \sigma^2 + \mathbf{p}^T \tilde{\mathbf{h}}_k \Big)^2 - 2b_k [\tilde{\mathbf{h}}_k]_j \Big( \sigma^2 + \mathbf{p}^T \tilde{\mathbf{h}}_e \Big)^2. \tag{35}$$

Thus, in order to prove $\boldsymbol{\psi}(\mathbf{p}) < \mathbf{0}$, it is equivalent to prove

$$\sqrt{c_k [\tilde{\mathbf{h}}_e]_j} \Big( \sigma^2 + \mathbf{p}^T \tilde{\mathbf{h}}_k \Big) < \sqrt{b_k [\tilde{\mathbf{h}}_k]_j} \Big( \sigma^2 + \mathbf{p}^T \tilde{\mathbf{h}}_e \Big). \tag{36}$$

or,

$$\left( \sqrt{c_k [\tilde{\mathbf{h}}_e]_j} - \sqrt{b_k [\tilde{\mathbf{h}}_k]_j} \right) \sigma^2 + \mathbf{p}^T \left( \sqrt{c_k [\tilde{\mathbf{h}}_e]_j} \tilde{\mathbf{h}}_k - \sqrt{b_k [\tilde{\mathbf{h}}_k]_j} \tilde{\mathbf{h}}_e \right) < 0. \tag{37}$$

Under the Conditions 2 and 3, we find that $\sqrt{c_k [\tilde{\mathbf{h}}_e]_j} - \sqrt{b_k [\tilde{\mathbf{h}}_k]_j} < 0$ and $\sqrt{c_k [\tilde{\mathbf{h}}_e]_j} \tilde{\mathbf{h}}_k - \sqrt{b_k [\tilde{\mathbf{h}}_k]_j} \tilde{\mathbf{h}}_e <$ 0, respectively. Therefore, the inequality in (37) is satisfied and the monotonicity is shown.

### C. Proof of Scalability

The scalability condition can be rewritten as (if $\rho > 1$)

$$\frac{\rho a_k}{\frac{b_k}{\sigma^2 + \mathbf{p}^T \tilde{\mathbf{h}}_k} - \frac{c_k}{\sigma^2 + \mathbf{p}^T \tilde{\mathbf{h}}_e}} > \frac{\rho a_k}{\frac{\rho b_k}{\sigma^2 + \rho \mathbf{p}^T \tilde{\mathbf{h}}_k} - \frac{\rho c_k}{\sigma^2 + \rho \mathbf{p}^T \tilde{\mathbf{h}}_e}}, \tag{38}$$

since $I_k(\mathbf{p}) \geq 0$, the condition in (38) is equivalent to

$$\frac{b_k}{\sigma^2 + \mathbf{p}^T \tilde{\mathbf{h}}_k} - \frac{c_k}{\sigma^2 + \mathbf{p}^T \tilde{\mathbf{h}}_e} < \frac{\rho b_k}{\sigma^2 + \rho \mathbf{p}^T \tilde{\mathbf{h}}_k} - \frac{\rho c_k}{\sigma^2 + \rho \mathbf{p}^T \tilde{\mathbf{h}}_e}. \tag{39}$$

Inequality (39) is equivalent to

$$\frac{\Delta}{(\sigma^2 + \mathbf{p}^T \tilde{\mathbf{h}}_k)(\sigma^2 + \mathbf{p}^T \tilde{\mathbf{h}}_e)(\sigma^2 + \rho \mathbf{p}^T \tilde{\mathbf{h}}_k)(\sigma^2 + \rho \mathbf{p}^T \tilde{\mathbf{h}}_e)} < 0, \tag{40}$$

where $\Delta$ is given by

$$\Delta = \sigma^6 \left( 1 - \rho \right) \left( b_k - c_k \right) + \sigma^4 \left( 1 - \rho^2 \right) \left( b_k \mathbf{p}^T \tilde{\mathbf{h}}_e - c_k \mathbf{p}^T \tilde{\mathbf{h}}_k \right) + \sigma^2 \rho \left( 1 - \rho \right) \left[ b_k (\mathbf{p}^T \tilde{\mathbf{h}}_e)^2 - c_k (\mathbf{p}^T \tilde{\mathbf{h}}_k)^2 \right], \tag{41}$$

where the condition in (40) $\Leftrightarrow$ "$\Delta < 0$". $b_k > c_k$ is satisfied under the Condition 1, and $b_k \mathbf{p}^T \tilde{\mathbf{h}}_e > c_k \mathbf{p}^T \tilde{\mathbf{h}}_k$, and $b_k (\mathbf{p}^T \tilde{\mathbf{h}}_e)^2 > c_k (\mathbf{p}^T \tilde{\mathbf{h}}_k)^2$ are satisfied under the Condition 2. Since $\rho > 1$, $\Delta$ in (41) is proved that $\Delta < 0$. Therefore, the scalability is also proved.

## Appendix B

## Proof of Theorem 2

As we proved in Appendix A, by replacing $\tilde{\mathbf{h}}_k$ and $\tilde{\mathbf{h}}_e$ with $\tilde{\mathbf{h}}_k = \mathbf{0}$, and

$$[\tilde{\mathbf{h}}_e]_j = \begin{cases} \mathbf{w}_j^H \widehat{\mathbf{R}}_e \mathbf{w}_j, & \text{if } j \neq k, \\ 0, & \text{otherwise,} \end{cases} \tag{42}$$

respectively, we will prove the positivity, monotonicity and scalability in the following.

### A. Proof of Positivity

$f(\mathbf{p})$ in (32) can be re-formulated as

$$f(\mathbf{p}) = \frac{b_k}{\sigma^2} - \frac{c_k}{\sigma^2 + \mathbf{p}^T \tilde{\mathbf{h}}_e} = \frac{\sigma^2 (b_k - c_k) + b_k \mathbf{p}^T \tilde{\mathbf{h}}_e}{\sigma^2 (\sigma^2 + \mathbf{p}^T \tilde{\mathbf{h}}_e)}. \tag{43}$$

Since $b_k \geq c_k$, it follows that $f(\mathbf{p}) > 0$, the positivity of $I_k(\mathbf{p})$ is proved.

### B. Proof of Monotonicity

$\boldsymbol{\varphi}(\mathbf{p})$ in (33) can be re-formulated with $\tilde{\mathbf{h}}_k = \mathbf{0}$ as

$$\boldsymbol{\varphi}(\mathbf{p}) = \frac{\partial f(\mathbf{p})}{\partial \mathbf{p}} = \frac{c_k \tilde{\mathbf{h}}_e}{\left(\sigma^2 + \mathbf{p}^T \tilde{\mathbf{h}}_e\right)^2} > 0. \tag{44}$$

Therefore, $f(\mathbf{p})$ increase monotonically with $\mathbf{p}$, the monotonicity of $I_k(\mathbf{p})$ is proved.

### C. Proof of Scalability

We can re-formulate $\Delta$ in (41) as (let $\tilde{\mathbf{h}}_k = \mathbf{0}$)

$$\Delta = \sigma^6 (1 - \rho)(b_k - c_k) + \sigma^4 \left(1 - \rho^2\right) b_k \mathbf{p}^T \tilde{\mathbf{h}}_e + \sigma^2 \rho (1 - \rho) b_k (\mathbf{p}^T \tilde{\mathbf{h}}_e)^2. \tag{45}$$

Since $\rho > 1$ and $b_k > c_k$ , $\Delta$ in (45) is shown that $\Delta < 0$. Therefore, the scalability is also proved.

### Acknowledgment

This work is partially supported by US NSF CNS-0953377, CNS-0905556, CNS-0910461, ECCS-1028782, and Research Council of Norway through projects 197565/V30, 183311/S10, and 176773/S10.

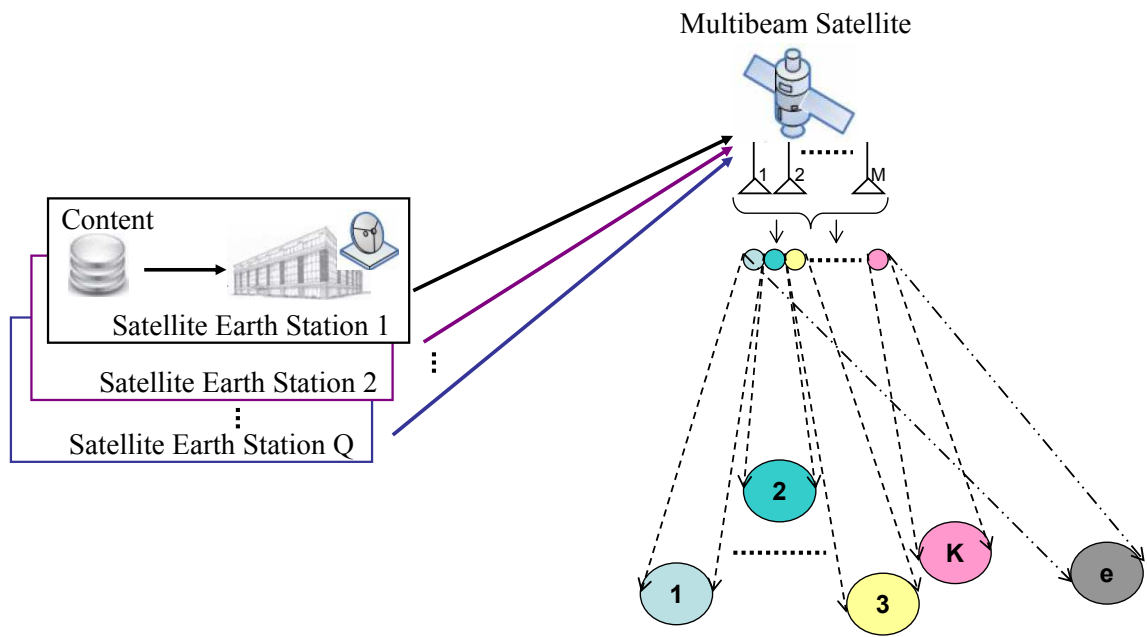

Fig. 1.  Multibeam SATCOM scenario.

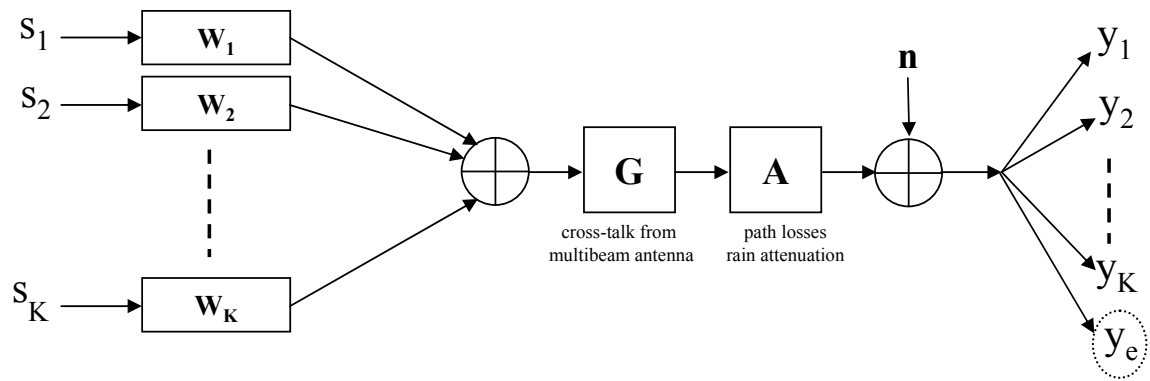

Fig. 2.   Block matrix model of the satellite broadcast channel.

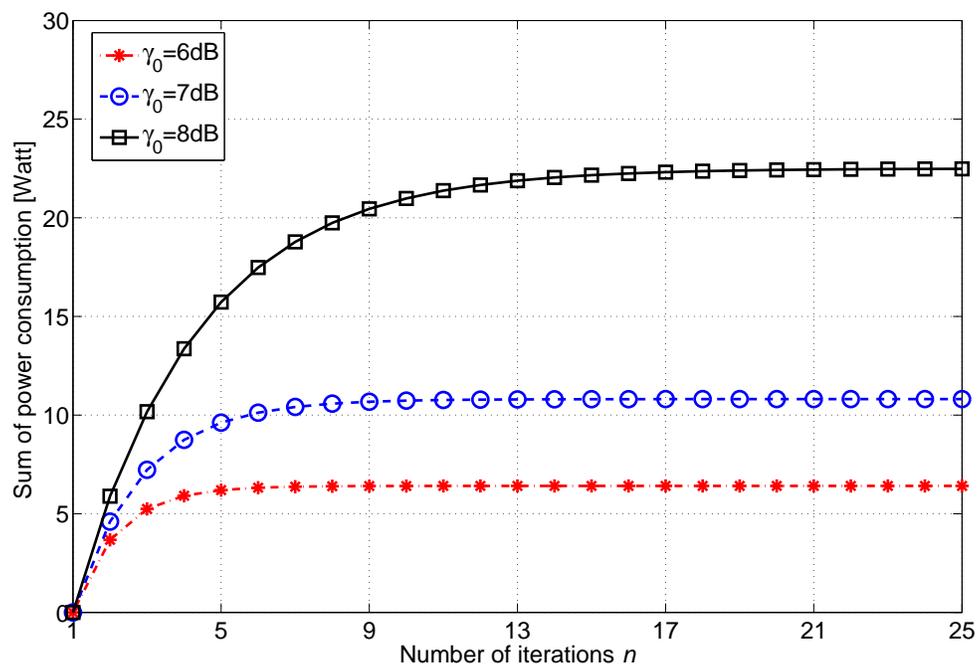

Fig. 3. Total transmitted power versus the iteration number with $M = 8$, $K = 5$, and $\alpha_k = \alpha_e = 0.8$ for $k = 1, 2, \ldots, 5$.

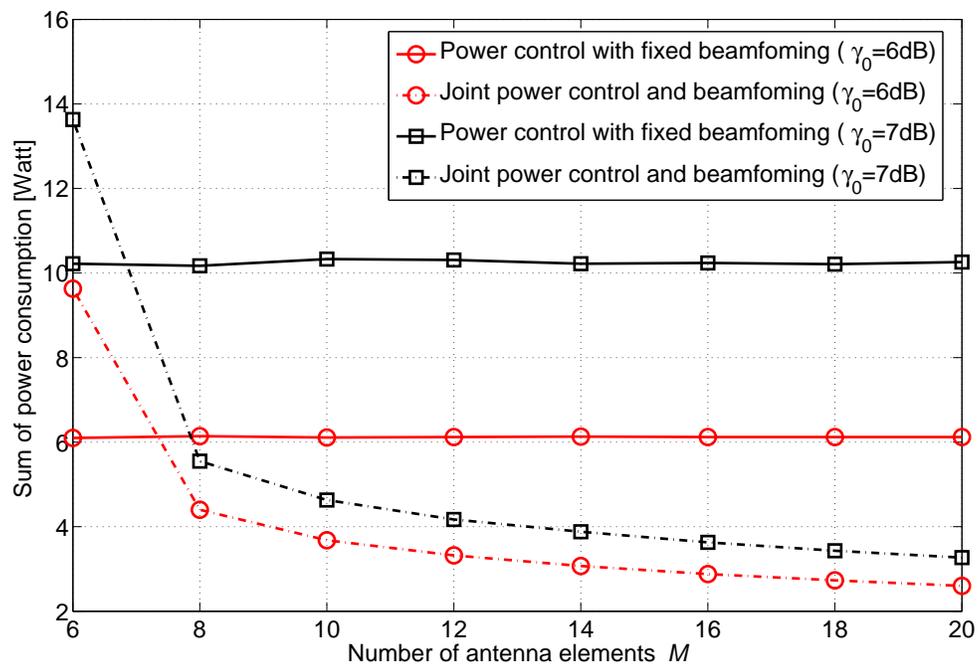

Fig. 4. Total transmitted power versus the number of antenna elements with $K = 5$ and $\alpha_k = \alpha_e = 0.8$ for $k = 1, 2, \dots, 5$.

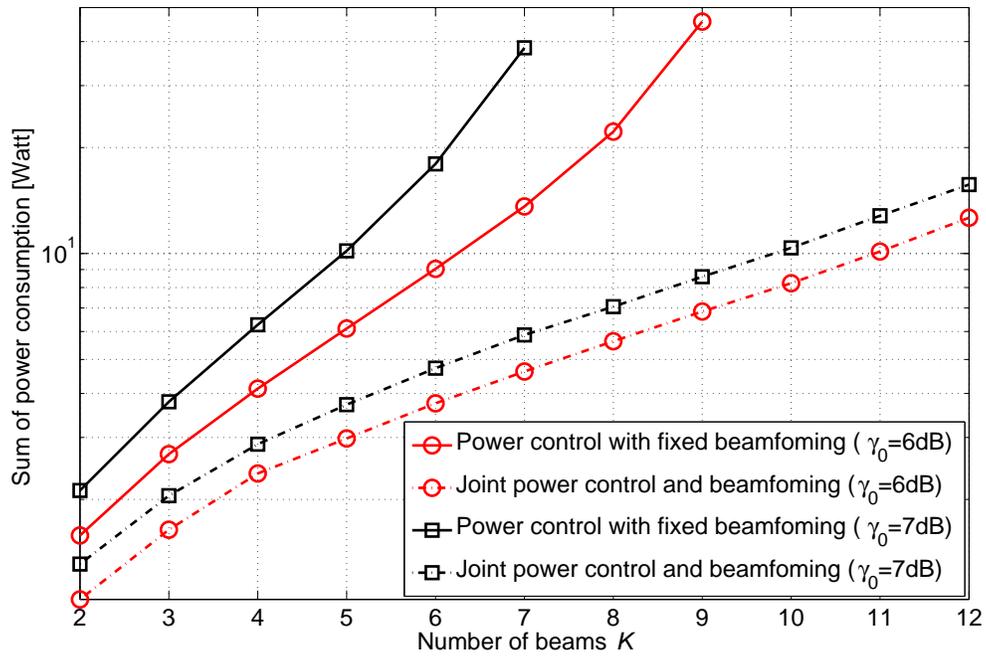

Fig. 5. Total transmitted power versus the number of beams with $M = 15$ and $\alpha_k = \alpha_e = 0.8$ for $k = 1, 2, \ldots, K$.

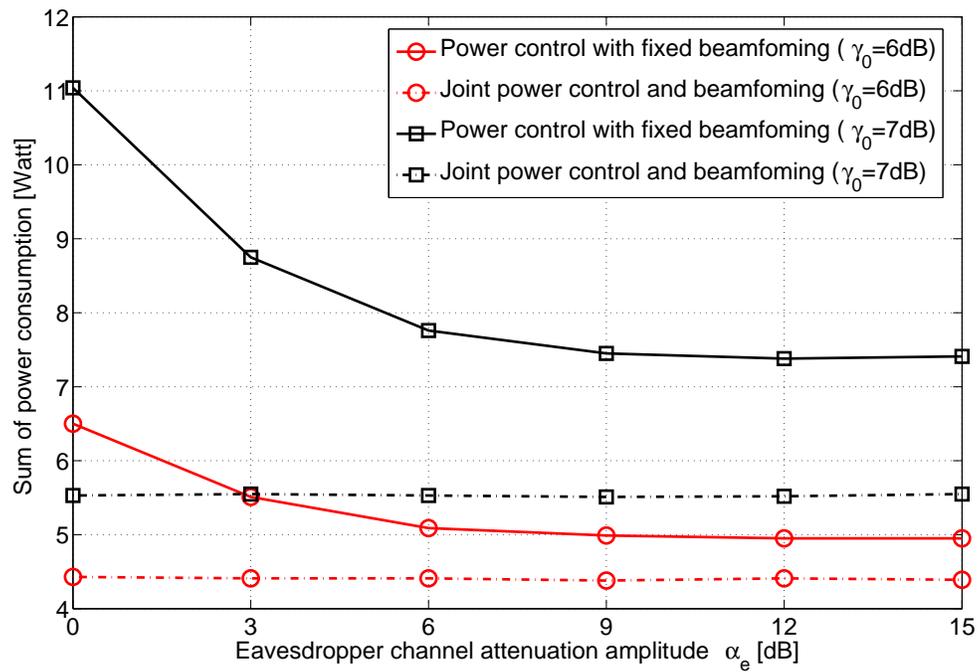

Fig. 6. Total power consumption versus the channel attenuation amplitude to the eavesdropper with $M = 8$, $K = 5$, and $\alpha_k = 0.8$ for $k = 1, 2, \ldots, 5$.

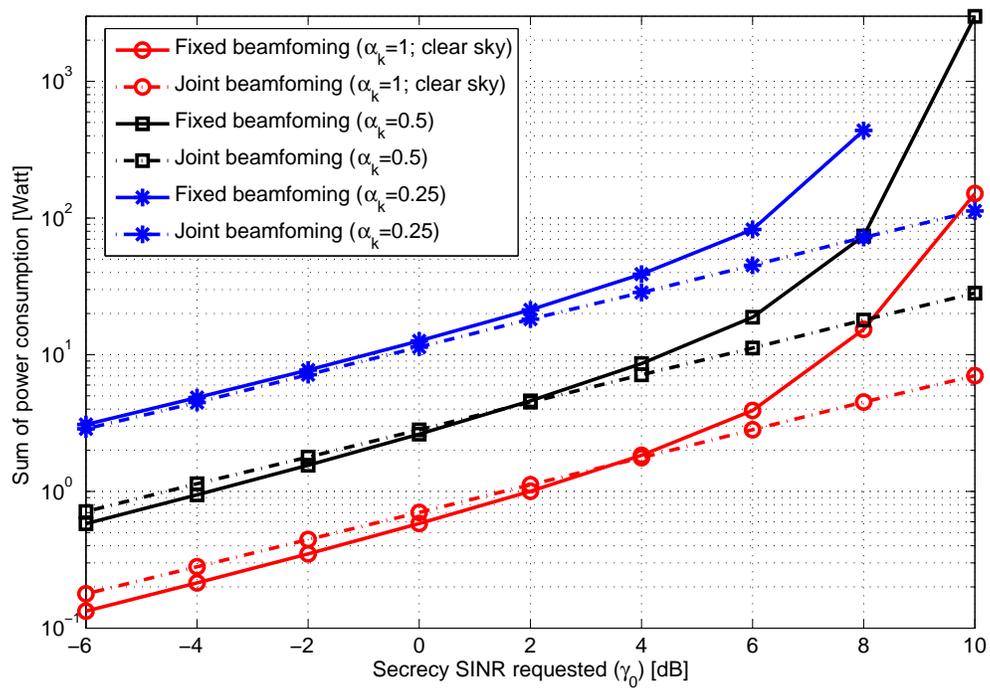

Fig. 7. Total transmitted power versus the target secrecy SINR with $M = 8$, $K = 5$, and $\alpha_e = 1$.

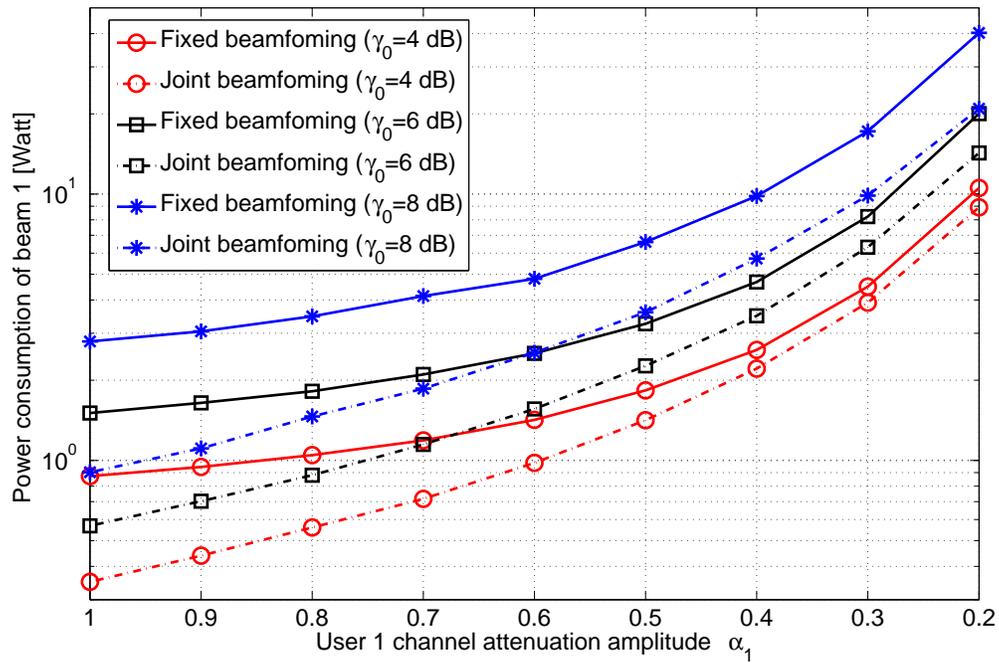

Fig. 8. Transmitted power for a specific beam (e.g., Beam 1) versus the channel condition with $M = 8$, $K = 5$, and $\alpha_k = \alpha_e = 0.8$ for $k = 2, 3, 4,$ and 5.

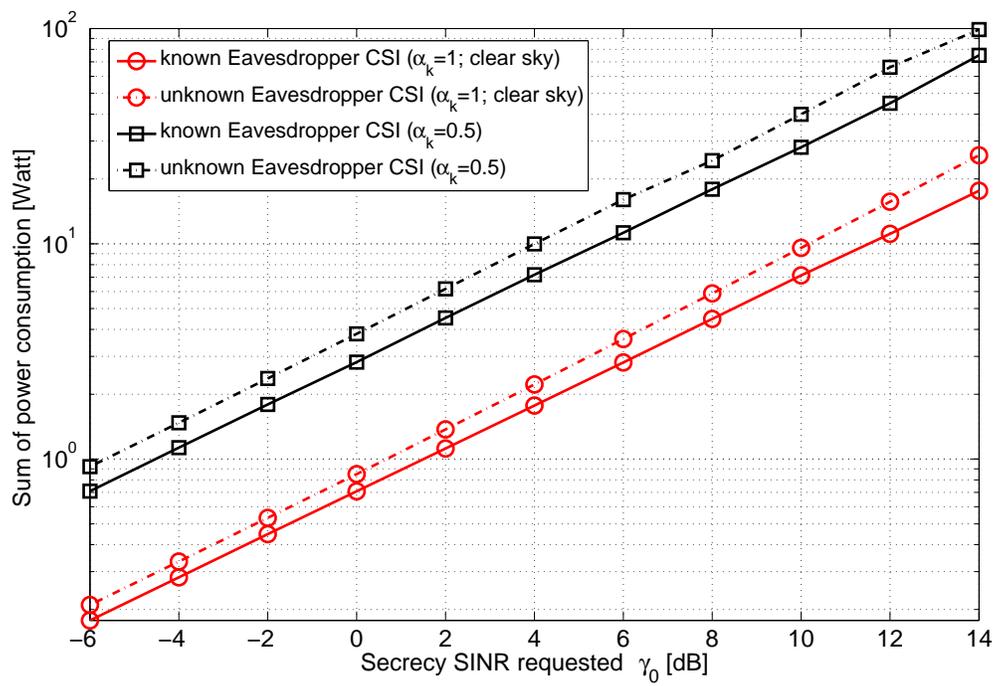

Fig. 9. Power allocation with or without the available the eavesdropper CSI with $M = 8$, $K = 5$ and $\alpha_e = 1$.

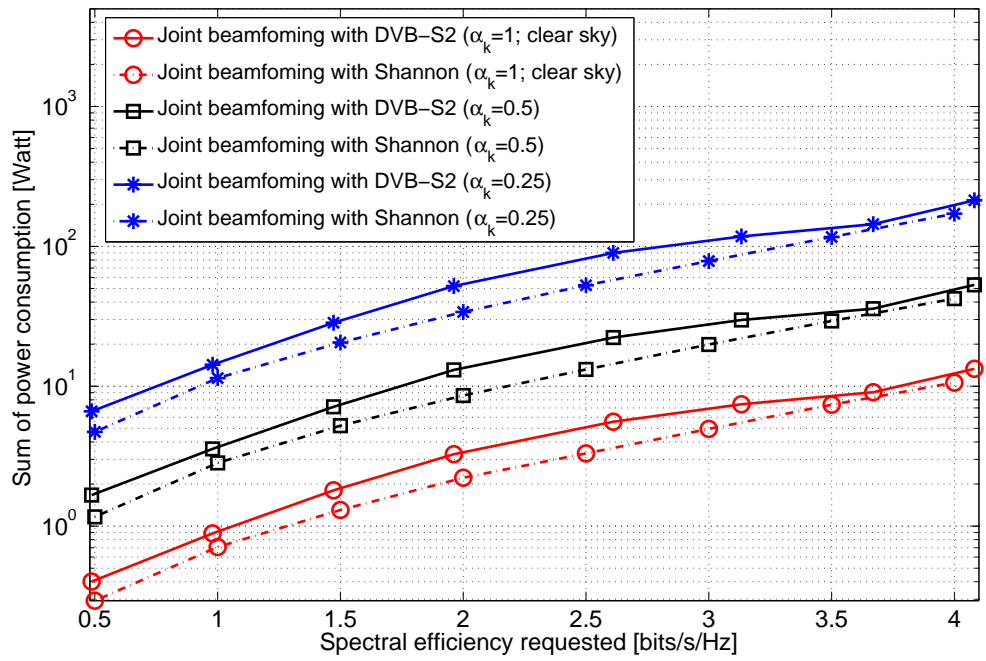

Fig. 10. Total transmitted power comparison for the DVB-S2 air-interface and Gaussian inputs with $M = 8$, $K = 5$, and $\alpha_e = 1$.

TABLE I

MAXIMUM NUMBER OF USERS ($P_{\text{TOT}} = 10$ WATT, $\gamma_0 = 6dB$, $M = 20$)

| System setup | Maximum number of users |
|---|---|
| Fixed power, fixed beamforming $\alpha_k = \alpha_e = 1$ | 4 |
| Power control, fixed beamforming $\alpha_k = \alpha_e = 1$ | 9 |
| Power control, fixed beamforming $\alpha_k = 1$, $\alpha_e = 0.5$ | 13 |
| Joint power control and beamforming $\alpha_k = \alpha_e = 1$ | 20 |
| Joint power control and beamforming $\alpha_k = 1$, $\alpha_e = 0.5$ | 21 |